\documentclass[%
reprint,
 amsmath,amssymb,
 aps,
]{revtex4-1}

\usepackage{graphicx}
\usepackage{dcolumn}
\usepackage{bm}
\usepackage{mhchem}
\usepackage{color}
\usepackage{dcolumn}
\usepackage[T1]{fontenc}

\begin{document}

\preprint{APS/123-QED}

\title{New Gd-based magnetic compound GdPt$_2$B with a chiral crystal structure}

\author{Yoshiki J. Sato$^{1,2}$}
\email{yoshiki_sato@rs.tus.ac.jp}
\author{Hikari Manako$^{1}$}
\author{Yoshiya Homma$^{2}$}
\author{Dexin Li$^{2}$}
\author{Ryuji Okazaki$^{1}$}
\author{Dai Aoki$^{2}$}
\affiliation{%
 $^{1}$Department of Physics, Faculty of Science and Technology, Tokyo University of Science, Noda, Chiba 278-8510, Japan\\
 $^{2}$Institute for Materials Research, Tohoku University, Oarai, Ibaraki 311-1313, Japan\\
}%


\date{\today}

\begin{abstract}
Herein, we report the discovery of a novel Gd-based magnetic compound GdPt$_2$B with a chiral crystal structure. X-ray diffraction and chemical composition analyses reveal a CePt$_2$B-type crystal structure (space group: $P$6$_4$22) for GdPt$_2$B. Moreover, we successfully grew single crystals of GdPt$_2$B using the Czochralski method. Magnetization measurements and the Curie--Weiss analysis demonstrate that the ferromagnetic interaction is dominant in GdPt$_2$B. A clear transition is observed in the temperature dependence of electrical resistivity, magnetic susceptibility, and specific heat at $T_{\rm O}$ = 87 K. The magnetic phase diagram of GdPt$_2$B, which consists of a field-polarized ferromagnetic region and a magnetically ordered region, resembles those of known chiral helimagnets. Furthermore, magnetic susceptibility measurements reveal a possible spin reorientation within the magnetically ordered phase in magnetic fields perpendicular to the screw axis. The results demonstrate that GdPt$_2$B is a suitable platform for investigating the competing effects of ferromagnetic and antisymmetric exchange interactions in rare-earth-based chiral compounds.
\end{abstract}

\maketitle
\section{INTRODUCTION}
Novel spin textures in inorganic chiral materials, such as magnetic skyrmion \cite{SM09,XZY10} and chiral soliton lattice \cite{YT12,SO14,TM17}, have received considerable attention because of their applicability to novel electronic devices. Generally, these characteristic spin textures are closely related to the symmetry of the crystal structure. In noncentrosymmetric crystal structures, an antisymmetric exchange interaction--the Dzyaloshinskii--Moriya (DM) interaction \cite{ID58,TM60}--affects the magnetic ground state. Moreover, competition between the exchange and DM interactions occasionally induces characteristic spin textures in chiral crystals.
 
In this paper, we report the discovery of a novel Gd-based compound GdPt$_2$B with a chiral crystal structure. For the 4$f^7$ configuration of trivalent Gd ion with a half-filled 4$f$ shell, Hund's rule yields $S$ = 7/2 and $L$ = 0. The quenched orbital momentum results in relatively weak spin-orbit coupling in trivalent Gd (and  divalent Eu) compounds and may lead to interesting spin textures, as observed in 3$d$ magnets. For instance, the anomalous Hall effect induced by topological spin textures has been reported in Gd/Eu-based frustrated magnets \cite{TK19,MH19,TS21} and Eu-based chiral helimagnets \cite{MK18,YO20}. However, Gd-based intermetallic compounds with chiral crystal structures are rarely reported. Therefore, GdPt$_2$B can provide a useful platform for exploring interesting spin textures induced by the antisymmetric exchange interaction.

\begin{figure*}[t]
\begin{center}
\includegraphics[width=\linewidth]{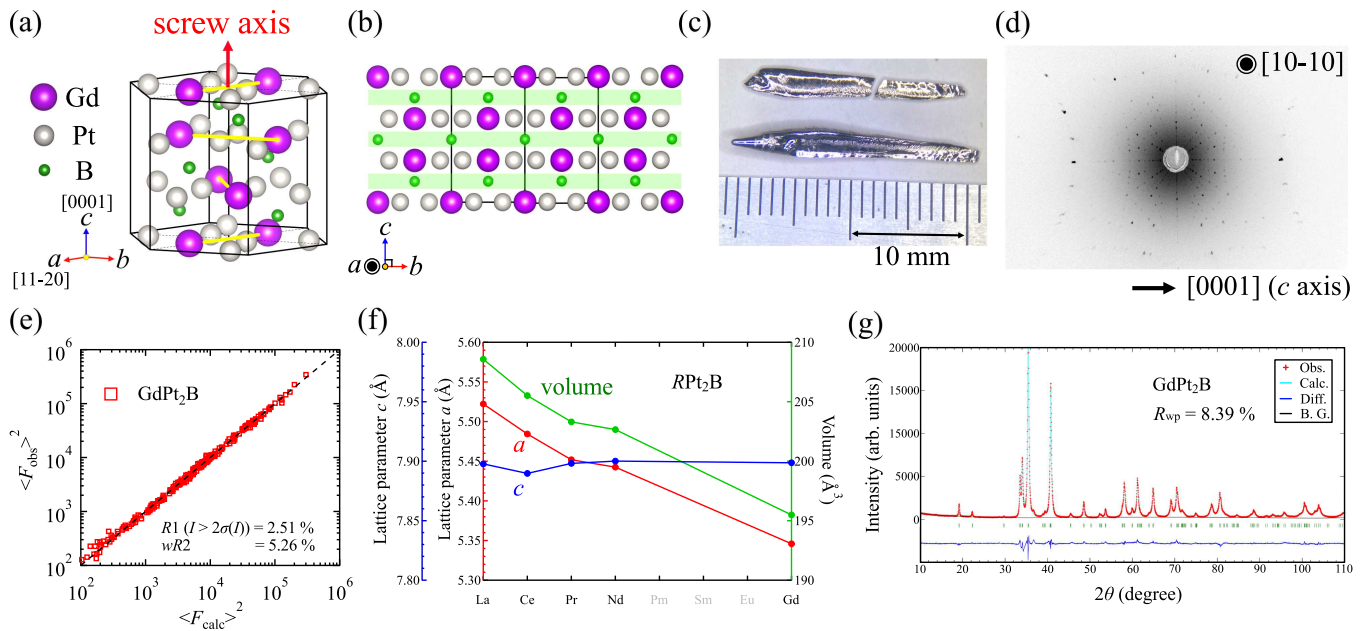}
\caption{\label{fig1} (a) Chiral crystal structure of left-handed (Space group: $P$6$_4$22) GdPt$_2$B.(b) Side view of the crystal structure. Images of crystal structures were constructed using the VESTA\cite{KM11}. (c) Single crystals of GdPt$_2$B grown using the Czochralski method. (d) Laue photograph of a GdPt$_2$B single crystal. (e) Plot of observed squared structure factors $\langle F_{\rm obs}\rangle^2$ of GdPt$_2$B single crystal as a function of calculated squared structure factors $\langle F_{\rm calc}\rangle^2$ on a logarithmic scale. (f) Plot of variations of lattice parameters and volume of $R$Pt$_2$B compounds. (g) X-ray diffraction powder pattern of GdPt$_2$B single crystals. $R_{\rm wp}$ is the weighted profile $R$ factor.}
\end{center}
\end{figure*}

As discussed in Section$~$\ref{sec3}, X-ray diffraction revealed that GdPt$_2$B crystallizes in a CePt$_2$B-type hexagonal structure [space group: $P$6$_2$22 or $P$6$_4$22, see Figs.$~$\ref{fig1}(a) and \ref{fig1}(b)]. To date, a series of $R$Pt$_2$B compounds with the CePt$_2$B-type structure has been reported for Y \cite{MD07}, La \cite{OLS03}, Ce \cite{OS00}, Pr \cite{OLS03}, Nd \cite{OLS03,YJS21}, Tm \cite{RTK10}, Yb \cite{RTK15}, and Lu \cite{RTK15}. The $R$Pt$_2$B compounds have the same point group $D_6$ as the typical monoaxial chiral helimagnets CsCuCl$_3$ \cite{AFW47,KA80} and Cr$_{1/3}$NbS$_2$ \cite{TM82,TM83}. As shown in Fig.$~$\ref{fig1}(a), Gd ions form a helical arrangement along the $[0001]$ axis, which can be regarded as a screw axis. The Gd--Pt and boron layers are stacked alternately along the $[0001]$ axis, as shown in Fig.$~$\ref{fig1}(b). Previously, we succeeded in growing single crystals of NdPt$_2$B using the Czochralski method \cite{YJS21}. Therefore, in addition to the synthesis of the polycrystal, we grew single crystals of GdPt$_2$B using the Czochralski method.

Additionally, this study investigates the electrical transport, magnetic, and thermodynamic properties of GdPt$_2$B single crystals. GdPt$_2$B exhibits a phase transition at $T_{\rm O}$ = 87 K. The temperature dependence of magnetization indicates that the magnetic transition is neither a simple ferromagnetic nor an antiferromagnetic transition, whereas the Curie--Weiss analysis suggests that the ferromagnetic interaction is dominant in GdPt$_2$B. A magnetic phase diagram is constructed from magnetic and thermodynamic measurements, and the magnetic ordering state of GdPt$_2$B is examined.

\section{EXPERIMENTAL DETAILS}
GdPt$_2$B polycrystals were synthesized via arc-melting using a tetra arc furnace under an argon atmosphere (99.9999 \% purity). The starting materials include stoichiometric amounts of Gd (99.9 \% purity), Pt (99.95 \% purity), and B (99.9 \% purity). The ingot was flipped and remelted five times to ensure homogeneity. A part of the synthesized polycrystals was used in single crystal growth. Single crystals of GdPt$_2$B were grown using the Czochralski method. The single crystal was pulled from the melting ingot using a tungsten seed at a speed of 12 mm h$^{-1}$. The Czochralski process was performed in a tetra arc furnace under an argon atmosphere (99.9999 \% purity). The obtained single crystals are shown in Fig.$~$\ref{fig1}(c). Single crystals of GdPt$_2$B are stable in air. The single crystals were oriented using a Laue camera (Photonic Science Laue X-ray CCD camera) and cut using a spark cutter for subsequent measurements. The Laue photograph of the GdPt$_2$B single crystal is shown in Fig.$~$\ref{fig1}(d).

The crystal structure of GdPt$_2$B was determined using a single-crystal X-ray diffractometer (Rigaku XtaLAB mini II) with Mo $K\alpha$ radiation ($\lambda$ = 0.71073 $\mathrm{\mathring{A}}$). The measured GdPt$_2$B single crystals were mounted on a plastic fiber. The data were collected at 296 K using a $\omega-2\theta$ scan. After the single-crystal x-ray diffraction (XRD), the homogeneity and single-phase nature of the grown single crystal were confirmed via powder XRD (Rigaku RINT2500V) with Cu $K\alpha$ radiation ($\lambda$ = 1.5418 $\mathrm{\mathring{A}}$). The chemical composition and homogeneity of the obtained single crystals were investigated through the inductively coupled plasma atomic emission spectroscopy (ICP-AES). Two samples (labeled as sample 1 and sample 2) were obtained from different parts of the grown single crystal and analyzed. 

The electrical resistivity was measured using a four-probe DC method in a Gifford--McMahon refrigerator. Magnetization measurements were performed using a commercial SQUID magnetometer (QD MPMS) at temperatures down to 2 K and in magnetic fields up to 2 T. The specific heat was measured using a relaxation method in magnetic fields using a quantum design DynaCool physical properties measurement system (QD PPMS).

\section{RESULTS and DISCUSSION}
\label{sec3}
First, we discuss the crystal structure analysis of GdPt$_2$B using powder and single-crystal XRD. The crystal structure of GdPt$_2$B was determined through single-crystal XRD. The crystal structure was solved using SHELXT\cite{GMS15A} and refined using SHELXL software \cite{GMS15C}. The crystal data, structural refinement, and atomic parameters of the GdPt$_2$B single crystals are summarized in Tables \ref{t1} and \ref{t2}. 
\begin{table}[htbp]
\caption{\label{t1} Crystallographic and structural refinement data obtained from single-crystal XRD.}
\begin{ruledtabular}
\begin{tabular}{lc}
Empirical formula & \ce{GdPt2B}\\
Formula weight & 558.24\\
Crystal system &hexagonal\\
Space group & $P6_422$ (\#181)\\
$a$ ($\mathrm{\mathring{A}}$) & 5.3459(3) \\ 
$c$ ($\mathrm{\mathring{A}}$) & 7.8987(6) \\ 
Volume ($\mathrm{\mathring{A}}^3$) & 195.49(2) \\ 
Formula units per cell (Z) & 3 \\
Number of measured reflections (total) & 1766\\
Number of measured reflections (unique) & 257\\
Cut off angle (2$\theta_{\rm max}$) & 66.1$^\circ$\\
$R1$ ($I$ $>$ 2.00$\sigma$($I$)) & 0.0240\\
$R$ (All reflections) & 0.0251\\
$wR2$ (All reflections) & 0.0526\\
Goodness of fit & 1.096\\
Flack parameter & -0.038(18)\\
Max Shift/Error in Final Cycle & 0.000\\
\end{tabular}
\end{ruledtabular}
\end{table}

\begin{table}[htbp]
\caption{\label{t2}Atomic positions and displacement parameters of \ce{GdPt2B}.}
\begin{ruledtabular}
\begin{tabular}{l c c c c c}
\multicolumn{6}{c}{GdPt$_2$B}\\ \hline
Atom & Site & {\it x} & {\it y} & {\it z} & $B_{eq}$\\ \hline
Gd & 3$c$ & 1/2 & 0 & 0 & 0.268(19) \\
Pt & 6$i$ & 0.15166(9) & 0.30332(9) & 0 & 0.356(17) \\
B & 3$d$ & 1/2 & 0 & 1/2 & 0.7(4) \\
\end{tabular}
\end{ruledtabular}
\end{table}

GdPt$_2$B crystallizes in the CePt$_2$B-type crystal structure (hexagonal $P$6$_4$22 space group), which was determined using a direct method. To determine the unit cell, 831 reflections were acquired. The 2$\theta$ range for determining the unit cell ranged from 8.7$^\circ$ to 66.3$^\circ$. An empirical absorption correction and correction for secondary extinction were applied. The data were corrected for the Lorentz and polarization effects. The crystal structure was successfully refined using anisotropic displacement parameters, and the $R$-factors were sufficiently small: $R1$ = 2.4\% for $I$ $>$ 2.00$s(I)$ and $wR2$ = 5.26\%. The equivalent isotropic atomic displacement parameter is defined as $B_{\rm eq}$ = $\frac{8}{3} \pi^2 [U_{11}(aa^\ast)^2 + U_{22}(bb^\ast)^2 + U_{33}(cc^\ast)^2 + 2 U_{12}(aa^\ast bb^\ast)\cos c + 2U_{13}(aa^\ast cc^\ast)\cos b + 2U_{23}(bb^\ast cc^\ast)\cos a]$, where $U_{ij}$ is summarized in Table \ref{t3}.
\begin{table*}[htbp]
\caption{\label{t3}Anisotropic displacement parameters of \ce{GdPt2B}.}
\begin{ruledtabular}
\begin{tabular}{l c c c c c c}
\multicolumn{7}{c}{GdPt$_2$B}\\ \hline
Atom & $U_{11}$ & $U_{22}$  & $U_{33}$ & $U_{12}$ & $U_{13}$ & $U_{23}$\\ \hline
Gd & 0.0020(4) & 0.0021(3) & 0.0060(4) & 0.0010(2) & 0 & 0 \\
Pt & 0.0029(3) & 0.0029(3) & 0.0076(3) & 0.0013(2) & -0.00103(15) & -0.00103(15) \\
B & 0.013(9) & 0.004(10) & 0.007(10) & 0.002(5) & 0 & 0 \\
\end{tabular}
\end{ruledtabular}
\end{table*}
$a$, $b$, and $c$ are the real-space cell lengths. $a^\ast$, $b^\ast$, and $c^\ast$ denote the reciprocal-space cell lengths. The Flack parameter is used to determine the absolute structure of the noncentrosymmetric crystal structure. The obtained Flack parameter was nearly zero, indicating that the present absolute structure ($P$6$_4$22) is correct. Figure$~$\ref{fig1}(e) shows the observed squared structure factors $\langle F_{\rm obs}\rangle^2$ as a function of calculated structure factors $\langle F_{\rm calc}\rangle^2$ using space group $P$6$_4$22.

Figure$~$\ref{fig1}(f) shows a comparison of the lattice parameters ($a$ and $c$) and volume $V$ of the $R$Pt$_2$B ($R$ = La, Ce, Pr, Nd, and Gd) compounds. The lattice parameters of LaPt$_2$B have been reported in Ref.$~$\cite{OLS03}, and other data were obtained through single-crystal XRD. The lattice parameter ($a$) and the volume ($V$) monotonically decrease. Comparing GdPt$_2$B with LaPt$_2$B, $a$ and $V$ decreased by 3.2\% and 6\%, respectively. The decrease in the unit cell volume is known as lanthanoid contraction. In contrast to the cases of $a$ and $V$, the lattice parameter $c$ did not change significantly, and the degree of the change was less than 0.1\%. Anisotropic changes in the lattice parameters of $R$Pt$_2$B were unusual. The anisotropic change in the lattice parameters of $R$Pt$_2$B has also been reported by powder XRD measurements of polycrystalline samples \cite{RTK15}. In another example of rare-earth platinum boride $R$Pt$_3$B, the changes in $a$ and $c$ were isotropic in the tetragonal structure \cite{OLS03_2}. The anisotropic change in the lattice parameters of $R$Pt$_2$B indicates a characteristic metallic bonding in this system.

The powder XRD patterns of crushed single crystals of GdPt$_2$B are shown in Fig.$~$\ref{fig1}(g). The observed peaks are well indexed to the CePt$_2$B-type structure, which is consistent with the crystal structure determined through the single-crystal XRD. Rietveld analysis was performed using the RIETAN-FP\cite{FI07}. The weighted profile and the unweighted profile $R$-factors were obtained to be $R_{\rm wp}$ = 8.39\% and $R_{\rm p}$ = 6.21\%, respectively. The lattice parameters obtained by the Rietveld analysis were consistent with those determined by the single-crystal XRD. Note that we also performed powder XRD measurements on the as-cast polycrystalline samples and obtained a single-phase XRD pattern. This indicates that GdPt$_2$B is a congruently melting compound.

In addition to powder and single-crystal XRD, we examined the chemical composition of the obtained single crystal through ICP-AES. The chemical compositions of sample1 and sample 2 are listed in Table$~$\ref{t4}. The chemical composition ratio was approximately Gd:Pt:B = 1:2:1, and no sample dependence of the chemical composition was observed. The results of the chemical composition analysis demonstrated the stoichiometric composition and the uniformity of the composition of GdPt$_2$B single crystals, even considering the errors associated with the spectroscopy and the preparation of sample solutions.
\begin{table}[htbp]
\caption{\label{t4} Chemical composition obtained by ICP-AES analysis.}
\begin{ruledtabular}
\begin{tabular}{l c c c }
\multicolumn{4}{c}{GdPt$_2$B single crystal}\\ \hline
Sample No. (meas.) & Gd (at. \%) & Pt (at. \%) & B (at. \%) \\ \hline
Sample1 (1st) & 25.4 & 49.4 & 25.2  \\
Sample1 (2nd) & 25.5 & 49.5 & 25.0 \\
Sample2 (1st) & 25.4 & 49.4 & 25.2 \\
Sample2 (2nd) & 25.5 & 49.4 & 25.1 \\
\end{tabular}
\end{ruledtabular}
\end{table}

\begin{figure}[htbp]
\begin{center}
\includegraphics[width=\linewidth]{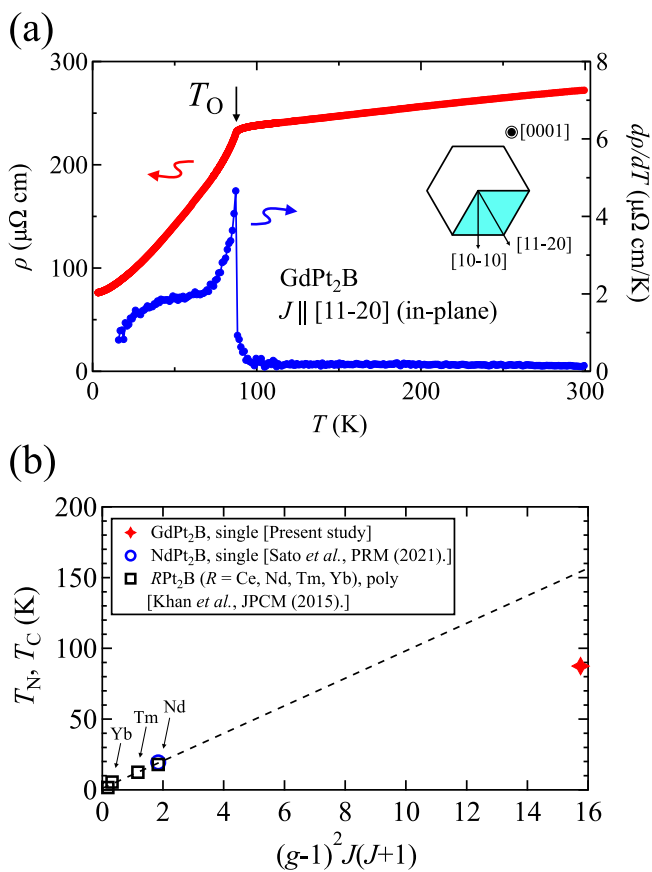}
\caption{\label{fig2} (a) Temperature dependence of the electrical resistivity $\rho(T)$ (red, left axis) and its temperature derivative $d\rho$/$dT$ (blue, right axis) of GdPt$_2$B for the electrical current $J$ || $[11\bar{2}0]$. (b) Plot of the ordering temperatures of $R$Pt$_2$B as a function of de Gennes factor $(g-1)^2J(J+1)$.}
\end{center}
\end{figure}

As we have shown, we succeeded in obtaining single-crystalline samples of GdPt$_2$B. Then, the basic physical properties of the GdPt$_2$B single crystals are presented. Figure$~$\ref{fig2}(a) shows the temperature dependence of the electrical resistivity $\rho$ of GdPt$_2$B for $J$ $||$ $[11\bar{2}0]$ (in-plane). 
In this study, we selected four-axis notation to indicate the crystallographic axes, as shown in the inset of Fig.$~$\ref{fig2}(a). GdPt$_2$B exhibited metallic behavior similar to that of other $R$Pt$_2$B systems. The ratio of the electrical resistivity at 300 K to that at the lowest temperature, namely the residual resistivity ratio, is approximately 3.5. $\rho(T)$ shows a clear anomaly at approximately $T_{\rm O}$ = 87 K. $d\rho$/$dT$ also shows a clear jump at $T_{\rm O}$.

The magnetism in metallic rare-earth compounds is expected to be based on the Ruderman--Kittel--Kasuya--Yosida (RKKY) interaction, and the ordering temperature could be proportional to the de Gennes factor $(g-1)^2J(J+1)$, where $g$ is the Land\'{e} $g$-factor. Khan$~${\it et~al.} examined the relationship between the transition temperature of $R$Pt$_2$B ($R$: Ce, Nd, Tm, Yb) and the de Gennes factor \cite{RTK15}. Here, we added the GdPt$_2$B data to the plot of the ordering temperatures of $R$Pt$_2$B as a function of the de Gennes factor, as shown in Fig.$~$\ref{fig2}(b). The transition temperature of GdPt$_2$B deviates from the reported scaling behavior. A similar deviation from the de Gennes scaling has been reported for $R$AuIn and $R$NiAl compounds, and the effect of the $d$-electrons of transition metal elements on the valence band spectra has also been reported in these systems \cite{LG08}. The energy band structure may affect the magnetism of the $R$Pt$_2$B series. Furthermore, the de Gennes scaling is often affected by the crystalline electric field (CEF). As observed in $R$PtGe compounds \cite{BP00}, the CEF effect may affect the transition temperatures of $R$Pt$_2$B compounds other than GdPt$_2$B.

\begin{figure}[t]
\begin{center}
\includegraphics[width=0.92\linewidth]{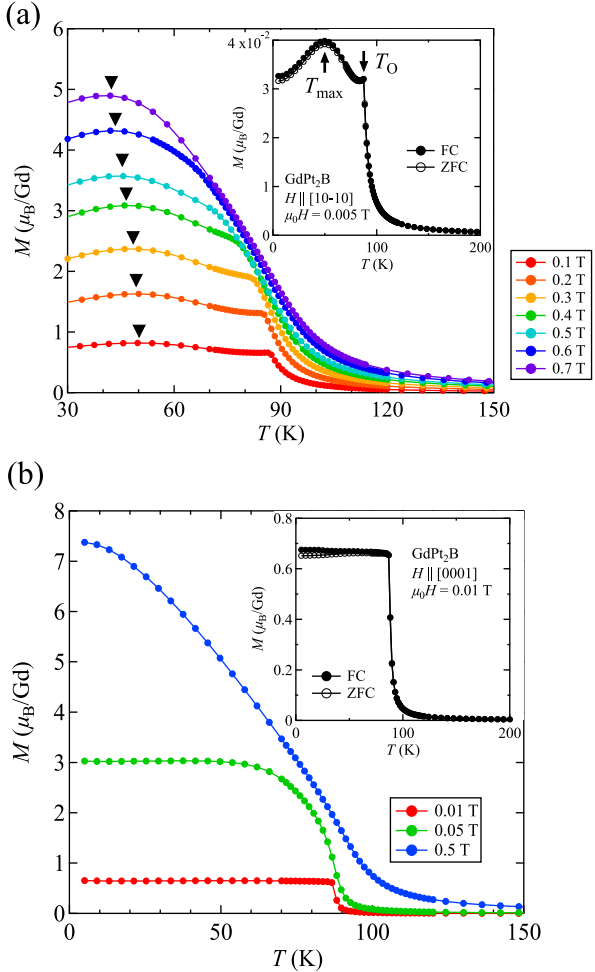}
\caption{\label{fig3} (a) Temperature dependence of the magnetization $M$ for $H$ $||$ $[10\bar{1}0]$ (in-plane) measured at various magnetic fields. The inset shows an enlarged view of $M$($T$) below the transition temperature $T_{\rm O}$ in the low magnetic field of 0.005 T. Both field-cooled (FC) and zero-field-cooled (ZFC) data are shown. (b) Temperature dependence of $M$ for $H$ $||$ $[0001]$ (out-of-plane) measured at 0.01, 0.05, and 0.5 T. The inset shows FC and ZFC data of $M$($T$) measured at 0.01 T. }
\end{center}
\end{figure}

To investigate the magnetic properties and transition at $T_{\rm O}$ of GdPt$_2$B, we measured the temperature dependence of the magnetization for $H$ $||$ $[10\bar{1}0]$ (in-plane) and $||$ $[0001]$ (out-of-plane). Figure \ref{fig3}(a) shows the temperature dependence of magnetization $M$ for $H$ $||$ $[10\bar{1}0]$ (in-plane).
As shown in the inset of Fig.$~$\ref{fig3}(a), $M(T)$ exhibits a rapid increase below approximately 100 K, followed by a peak structure at $T_{\rm O}$ in lower magnetic fields. This behavior indicates that the ground state of GdPt$_2$B is close to a ferromagnetic state but not a simple ferromagnetic ground state. A small bifurcation was observed between the field-cooled (FC) and zero-field-cooled (ZFC) curves below $T_{\rm O}$. Furthermore, $M(T)$ exhibits a broad maximum at $T_{\rm max}$ $\sim$ 50 K in both the FC and ZFC curves. The peak at $T_{\rm O}$ and the broad maximum at $T_{\rm max}$ shift to lower temperatures with increasing field.

In contrast to the temperature dependence of magnetization for $H$ $||$ $[10\bar{1}0]$ (in-plane), the out-of-plane $M(T)$ exhibits the nearly ferromagnetic behavior. The peak structure was not observed in the magnetization of $H$ $||$ $[0001]$. As shown in the inset of Fig.$~$\ref{fig3}(b), $M(T)$ increases abruptly below $T_{\rm O}$. Broad maximum of magnetization was absent for $H$ $||$ $[0001]$ in the ordered phase. The bifurcation between the FC and ZFC curves was also observed in out-of-plane $M(T)$ below $T_{\rm O}$, although the difference was smaller compared to that of the Gd-based ferromagnet\cite{SS17}. The sharp increase in the magnetization at $T_{\rm O}$ becomes broad in higher magnetic fields, similar to the case with ferromagnets, as shown in Fig.$~$\ref{fig3}(b). The characteristic anisotropic behavior of $M(T)$ indicates that the magnetic ground state of GdPt$_2$B is neither simple ferromagnetic nor antiferromagnetic.

\begin{figure}[t]
\begin{center}
\includegraphics[width=0.95\linewidth]{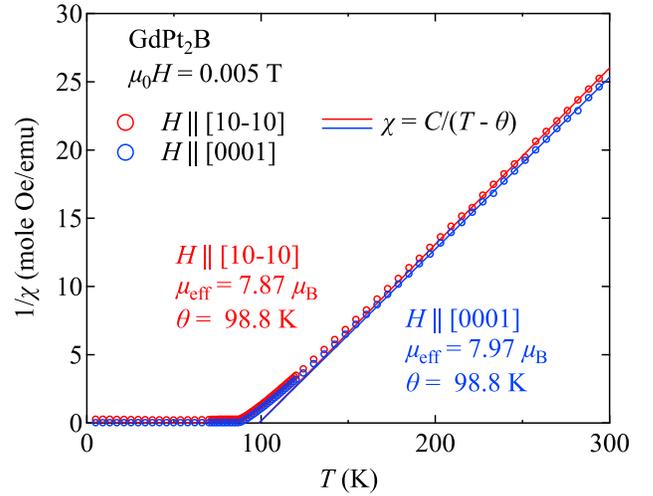}
\caption{\label{fig4} Temperature dependence of the inverse susceptibility 1/$\chi$ for $H$ $||$ $[10\bar{1}0]$ (red markers) and $[0001]$ (blue markers). The solid lines represent a least-squared fit obtained with the Curie-Weiss law above 200 K.}
\end{center}
\end{figure}

Figure \ref{fig4} shows the temperature dependence of the inverse susceptibility 1/$\chi$ $\equiv$ $H/M$ for $H$ $||$ $[10\bar{1}0]$ and $||$ $[0001]$. 
The anisotropy in the magnetic susceptibility is quite small in the paramagnetic region of GdPt$_2$B because of the configuration of the Gd$^{3+}$ ion ($S$ = 7/2, $L$ = 0). The small magnetocrystalline anisotropy in the paramagnetic state has been reported for other $S$ = 7/2 systems\cite{SN18,AM15}. GdPt$_2$B exhibits the Curie--Weiss behavior above 200 K. As shown in Fig.$~$\ref{fig4}, the magnetic susceptibility data are fitted to the expression of $\chi = C/(T-\theta)$, where $C$ and $\theta$ are the Curie constant and the Weiss temperature, respectively. We estimate the effective moment from the Curie--Weiss analysis as $\mu_{\rm eff}$ = 7.87 $\mu_{\rm B}$ per Gd for $H$ $||$ $[10\bar{1}0]$ and $\mu_{\rm eff}$ = 7.97 $\mu_{\rm B}$ per Gd for $||$ $[0001]$. The obtained $\mu_{\rm eff}$ is in reasonable agreement with the theoretical value of 7.94 $\mu_{\rm B}$ for free-ion Gd$^{3+}$ \cite{JJ91}. We obtained the Weiss temperature $\theta$ = 98.8 K for both principal axes, indicating that the interaction between the Gd-ions is ferromagnetic in GdPt$_2$B.

\begin{figure}[hptb]
\begin{center}
\includegraphics[width=0.95\linewidth]{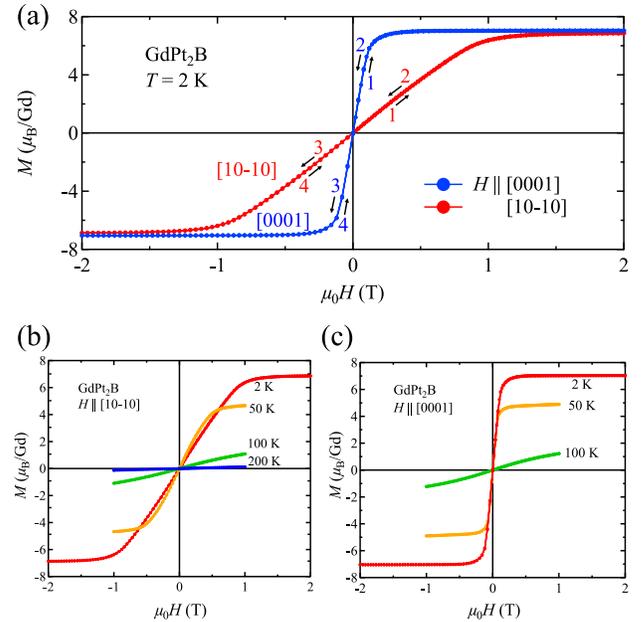}
\caption{\label{fig5} (a) Magnetization curves measured at 2 K for $H$ $||$ $[10\bar{1}0]$ (red markers) and $[0001]$ (blue markers). (b) $M(H)$ curves measured at several constant temperatures for $H$ $||$ $[10\bar{1}0]$. (c) $M(H)$ curves measured at several constant temperatures for $H$ $||$ $[0001]$.}
\end{center}
\end{figure}

As mentioned previously, $M(T)$ and $1/\chi(T)$ data demonstrate that the ferromagnetic interaction is dominant in GdPt$_2$B. Therefore, we measured the isothermal $M$($H$) curves to investigate the hysteresis loop behavior, as shown in Figs.$~$\ref{fig5}(a)--(c).
The measurement was performed by sweeping the magnetic field from $-2$ T to 2 T after zero-field cooling of the sample. The arrows in Fig$~$\ref{fig5}(a) indicate the measurement sequences. Despite the dominant ferromagnetic interaction, no signs of spontaneous magnetization were observed. The $M(H)$ curves change linearly with the sweeping magnetic field in the lower field region, regardless of the field direction. In contrast to the small magnetic anisotropy in the paramagnetic state, significant magnetic anisotropy was observed at temperatures below $T_{\rm O}$. The saturated magnetic moment was approximately 7 $\mu_{\rm B}$/Gd for both $[10\bar{1}0]$ and $[0001]$ axes. The magnetization saturates at approximately 0.5 T for $H$ $||$ $[0001]$, whereas $M(H)$ reaches 7 $\mu_{\rm B}$/Gd at 2 T for $H$ $||$ $[10\bar{1}0]$. As shown in Fig.$~$\ref{fig5} (b), the isothermal $M(H)$ curve measured at 50 K exceeded the $M(H)$ curve measured at the lowest temperature in the magnetic field range below 0.7 T for $H$ $||$ $[10\bar{1}0]$. This is consistent with the temperature-dependent measurements. Meanwhile, the isothermal magnetization curves measured at various constant temperatures exhibit a nearly ferromagnetic behavior for $H$ $||$ $[0001]$, as shown in Fig.$~$\ref{fig5} (c). Hysteresis loop behavior in $M(H)$ was absent for both principal crystallographic axes over the entire temperature range below the ordering temperature.

\begin{figure}[t]
\begin{center}
\includegraphics[width=0.95\linewidth]{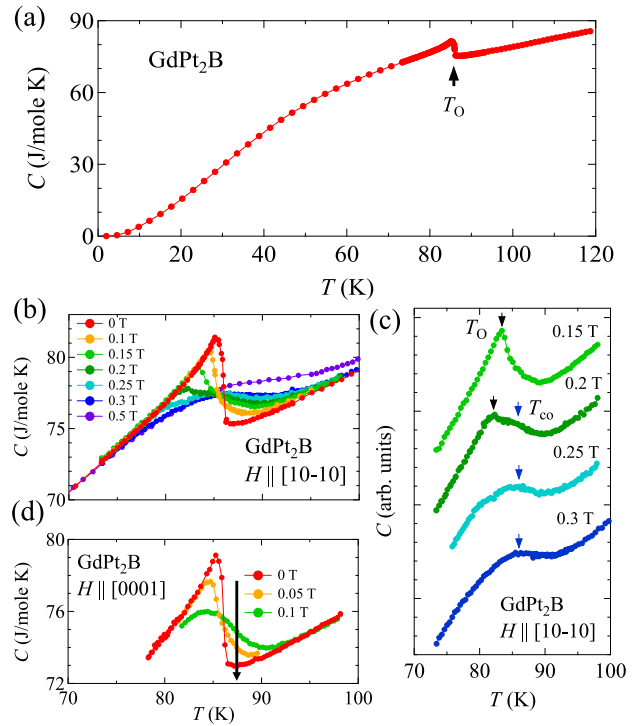}
\caption{\label{fig6} (a) Temperature dependence of the specific heat $C$ of GdPt$_2$B. (b) $C(T)$ measured in zero field and constant magnetic fields for $H$ $||$ $[10\bar{1}0]$. (c) $C(T)$ measured at 0.15, 0.2, 0.25, and 0.3 T. The specific heat data are shifted vertically for clarity. $T_{\rm O}$ and $T_{\rm co}$ are indicated by arrows. (d) $C(T)$ measured at 0, 0.05, and 0.1 T for $H$ $||$ $[0001]$}
\end{center}
\end{figure}

To clearly detect the phase transitions, we investigated the thermodynamic properties. Figure$~$\ref{fig6}(a) shows the temperature dependence of the specific heat $C$ of GdPt$_2$B. 
$C(T)$ exhibits a clear jump at $T_{\rm O} = 87$ K. Applying the magnetic field along the $[10\bar{1}0]$ axis, the anomaly shifts to lower temperatures in magnetic fields of up to 0.2 T, as shown in Figs.$~$\ref{fig6}(b) and \ref{fig6}(c). The field dependence of the anomaly in $C(T)$ at $T_{\rm O}$ is consistent with that of the peak structure in $M(T)$. In higher magnetic fields above 0.25 T, a broad maximum corresponding to the crossover between the paramagnetic and field-polarized ferromagnetic states was observed at $T_{\rm co}$. Notably, the anomaly at $T_{\rm O}$ coexists with the broad maximum at $T_{\rm co}$ in a magnetic field of 0.2 T, as shown in Fig.$~$\ref{fig6}(c). This behavior also indicates that the magnetically ordered phase is neither simple ferromagnetic nor antiferromagnetic phases.

$C(T)$ shows no apparent anomaly at $T_{\rm max}$ $\sim$ 50 K, where the magnetic susceptibility exhibits a broad maximum for $H$ $||$ $[10\bar{1}0]$. Note that the electrical resistivity also exhibited no clear anomalies at $T_{\rm max}$. The absence of an anomaly in the specific heat indicates that the broad maximum in the magnetic susceptibility at $T_{\rm max}$ is not due to a phase transition. The broad maximum at $T_{\rm max}$ suggests the occurrence of spin reorientation in the magnetically ordered phase. A similar anomaly in $M(T)$ derived from spin reorientation has been reported in several systems, such as Tb$_5$Si$_2$Ge$_2$ \cite{JPA05,NM19} and YBaCo$_4$O$_8$ \cite{AKB13,CD19}. Further studies using microscopic probes are required to confirm the origin of the broad anomaly in $M(T)$.

\begin{figure}[htpb]
\begin{center}
\includegraphics[width=0.95\linewidth]{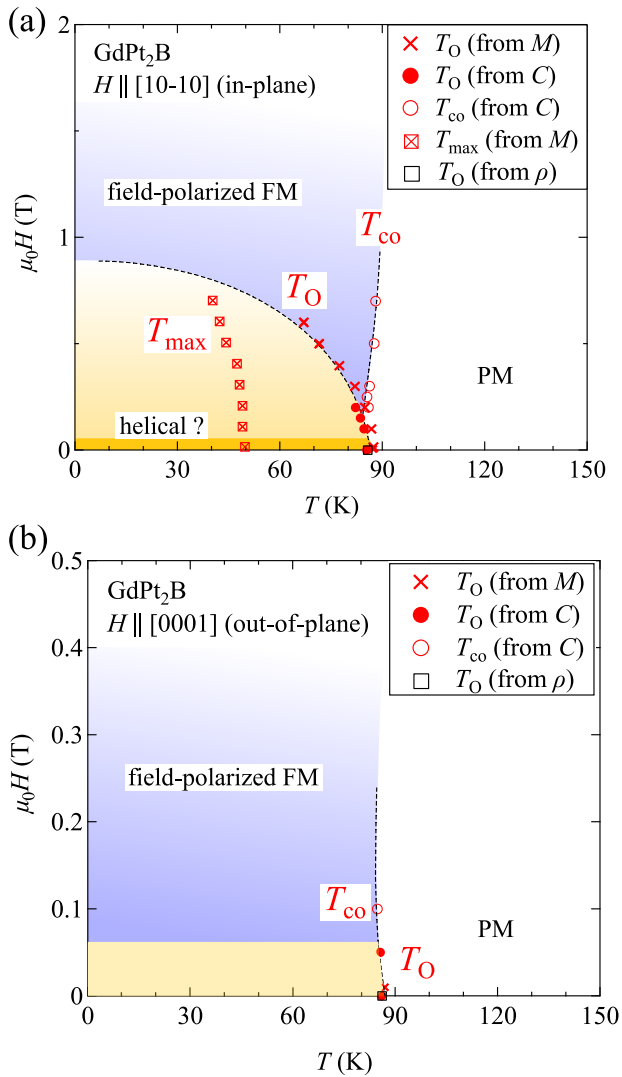}
\caption{\label{fig7} Magnetic phase diagram of GdPt$_2$B for (a) $H$ $||$ $[10\bar{1}0]$ ($H$ $\perp$ screw axis) and (b) $H$ $||$ $[0001]$ ($H$ $||$ screw axis). The dashed lines in the phase diagrams are guides to the eyes.}
\end{center}
\end{figure}

We constructed magnetic field--temperature ($H$--$T$) phase diagrams from the magnetization and thermodynamic measurements of $H$ $||$ $[10\bar{1}0]$ and $||$ $[0001]$, as shown in Figs.$~$\ref{fig7}(a) and (b). 
The $H$--$T$ phase diagram comprises field-polarized ferromagnetic and magnetically ordered regions below $T_{\rm O}$ under a magnetic field applied perpendicular to the screw axis (i.e., $H$ $||$ in-plane). The transition temperature $T_{\rm O}$ determined by magnetization measurements and that determined by specific heat measurements were consistent with each other. $T_{\rm O}$ shifts to lower temperatures under higher external magnetic fields. However, the crossover between the paramagnetic and field-polarized ferromagnetic states at $T_{\rm co}$ moves to higher temperatures with increasing magnetic fields. The in-plane $H$--$T$ phase diagram of GdPt$_2$B is reminiscent of known chiral helimagnets such as Cr$_{1/3}$NbS$_2$ \cite{NJG13,DAM22}, B20-type helimagnets \cite{ET17}, and rare-earth based chiral helimagnet YbNi$_3$Al$_9$ \cite{RM12,HN18}.

In contrast to the case of $H$ $||$ $[10\bar{1}0]$ ($H$ $\perp$ screw axis), the $H$--$T$ phase diagram for $H$ $||$ $[0001]$ ($H$ $||$ screw axis) is simple and resembles a ferromagnetic $H$--$T$ phase diagram. Fig. \ref{fig6}(d) shows the temperature dependence of the specific heat of GdPt$_2$B for $H$ $||$ $[0001]$. Similar to the behavior observed in the magnetization, the phase transition at $T_{\rm O}$ becomes crossover at magnetic fields as small as 0.1 T. In our magnetization and specific heat measurements, it was difficult to define the magnetically ordered region for $H$ $||$ $[0001]$. Electrical transport measurements in magnetic fields may be suitable for detecting phase boundaries, as reported for the monoaxial chiral helimagnet Cr$_{1/3}$NbS$_2$ \cite{YT13,YT15}.

Herein, we discuss the magnetically ordered state of GdPt$_2$B below $T_{\rm O}$. The temperature dependence of the magnetization exhibits a peak only for a magnetic field perpendicular to the screw axis. The peak structure in $M(T)$ and its anisotropy of GdPt$_2$B are similar to the magnetic behavior in the monoaxial chiral helimagnet Cr$_{1/3}$NbS$_2$ \cite{TM82,TM83} and molecule-based chiral magnet [Cr(CN)$_6$][Mn($S$)-pnH(H$_2$O)](H$_2$O)] \cite{JK05}. The formation of the peak structure in $M(T)$ of the Dzyaloshinskii--Moriya helimagnets is attributed to the commensurate-incommensurate (C-IC) transition \cite{AZ97,AZ98}. Kishine$~${\it et$~$al.} provided a theoretical interpretation of the peak formation in $M(T)$ by averaging the projection of the spin onto the field direction over the period of the chiral soliton lattice \cite{JK05}. The peak structure of $M(T)$ appears to occur due to the C-IC phase transition in GdPt$_2$B.

Note that no metamagnetic transition-like anomaly, which is often observed in helical magnets, was observed in the magnetization curve of GdPt$_2$B. In monoaxial chiral helimagnets, a steep increase in the magnetization associated with the formation of the chiral soliton lattice was observed under magnetic fields perpendicular to the screw axis \cite{TM82,TM83,SO14}. The magnetization process of GdPt$_2$B and its anisotropy are rather similar to those of a molecule-based chiral magnet K$_{0.4}$[Cr(CN)$_6$][Mn($S$)-pn]($S$)-pnH$_{0.6}$, which does not show a steep upturn in magnetization under magnetic fields perpendicular to the screw axis \cite{JK05}. The strength ratio between the DM and exchange interactions, namely $D/J$, plays a crucial role in forming the chiral soliton lattice under external fields. The $D/J$ ratio varies with the system, and whether clear formation of the chiral soliton lattice is observed could depend on the material. The absence of anomalies in the $M(H)$ curves of GdPt$_2$B could be related to the ratio of the strength of the exchange interaction to the strength of the DM interaction. Notably, the formation of chiral soliton lattice differs depending on the strength of the exchange interaction in the doped-YbNi$_3$Al$_9$ system \cite{TM17}. Moreover, the spin reorientation for $H$ $||$ $[10\bar{1}0]$ contributes to the magnetization process at temperatures below $T_{\rm max}$ $\sim$ 50 K. Further experiments, such as neutron scattering and/or resonant X-ray diffraction, are needed to reveal the detailed properties of the magnetically ordered state in GdPt$_2$B.

\section{Conclusions}
This paper reported the discovery of a new Gd-based magnetic compound GdPt$_2$B with the CePt$_2$B-type chiral crystal structure (space group: $P$6$_4$22, point group: $D_6^5$). The crystal structure and single-phase nature were analyzed by combining X-ray diffraction and chemical composition analysis. Additionally, we reported the growth of single crystals of GdPt$_2$B using the Czochralski method. The physical properties of GdPt$_2$B single crystals were examined by means of electrical resistivity, magnetization, and specific heat measurements. A clear phase transition was observed in $\rho(T)$, $M(T)$, and $C(T)$ at $T_{\rm O}$ = 87 K. Interestingly, $M(T)$ exhibited the characteristic peak structure at $T_{\rm O}$ for $H$ $\perp$ screw axis, whereas the peak structure was absent for $H$ $||$ screw axis.
We constructed  $H$--$T$ phase diagrams from the magnetization and thermodynamic measurements.
Based on the magnetization behavior and $H$--$T$ phase diagrams, GdPt2B could be a chiral helimagnet.
Our results demonstrate that GdPt$_2$B is a suitable platform for investigating the competing effects of ferromagnetic and antisymmetric exchange interactions in Gd-based chiral compounds.

\begin{acknowledgments}
We would like to thank Y. Haga, S. Ohara, F. Honda, Y. Shimizu, A. Nakamura for discussions.
We also thank F. Sakamoto (Analytical Research Core for Advanced Materials, Institute for materials Research, Tohoku University) for ICP-AES analysis.
We acknowledge all the support from the International Research Center for Nuclear Materials Science at Oarai (IMR, Tohoku University).
This work was supported by JSPS KAKENHI (JP22K20360, JP19H00646, JP20K20889), Grant-in-Aid for JSPS Research Fellow (JP19J20539), GIMRT Program (202205-IRKAC-0049), and DIARE research grant.
\end{acknowledgments}

\bibliography{GdPt2B}

\end{document}